\documentclass[11pt,preprintnumbers,titlepage,nofootinbib]{revtex4-2}
\usepackage[latin9]{inputenc}
\usepackage{amsmath}
\usepackage{amssymb}
\usepackage{graphicx}
\usepackage[unicode=true,
 bookmarks=false,
 breaklinks=false,pdfborder={0 0 1},backref=section,colorlinks=false]
 {hyperref}
\hypersetup{
 colorlinks,linkcolor=blue,citecolor=blue,urlcolor=blue}

\makeatletter

\usepackage{wasysym}
\usepackage{array}
\usepackage{multirow}
\usepackage{wasysym}
\usepackage{wasysym}
\usepackage{amsfonts}
\usepackage{subfigure}
\usepackage{cleveref}
\usepackage{listings}
\usepackage{xcolor}
\usepackage{array}
\usepackage{url}
\usepackage{color}

\@ifundefined{textcolor}{}{
\definecolor{BLACK}{gray}{0}
\definecolor{WHITE}{gray}{1}
\definecolor{RED}{rgb}{1,0,0}
\definecolor{GREEN}{rgb}{0,1,0}
\definecolor{BLUE}{rgb}{0,0,1}
\definecolor{CYAN}{cmyk}{1,0,0,0}
\definecolor{MAGENTA}{cmyk}{0,1,0,0}
\definecolor{YELLOW}{cmyk}{0,0,1,0}
}

\makeatother

\begin{document}
\preprint{CTP-SCU/2023018}
\title{Scalarized Kerr-Newman Black Holes}
\author{Guangzhou Guo$^{a,b}$}
\email{guogz@sustech.edu.cn}

\author{Peng Wang$^{a}$}
\email{pengw@scu.edu.cn}

\author{Houwen Wu$^{a,c}$}
\email{hw598@damtp.cam.ac.uk}

\author{Haitang Yang$^{a}$}
\email{hyanga@scu.edu.cn}

\affiliation{$^{a}$Center for Theoretical Physics, College of Physics, Sichuan
University, Chengdu, 610064, China}
\affiliation{$^{b}$Department of Physics, Southern University of Science and Technology,
Shenzhen, 518055, China}
\affiliation{$^{c}$Department of Applied Mathematics and Theoretical Physics,
University of Cambridge, Wilberforce Road, Cambridge, CB3 0WA, UK}
\begin{abstract}
In this paper, we construct scalarized rotating black holes within
the framework of Einstein-Maxwell-scalar models. These models incorporate
non-minimal couplings that can induce tachyonic instabilities, leading
to the spontaneous scalarization of Kerr-Newman (KN) black holes.
By exploring the domain of existence, we observe that the presence
of scalarized KN black holes is suppressed by the black hole spin,
with a maximum spin threshold beyond which scalarized solutions cease
to exist. Intriguingly, we find that in specific parameter regimes,
scalarized KN black holes can exhibit the presence of two unstable
and one stable light rings on the equatorial plane, manifesting in
both prograde and retrograde directions.
\end{abstract}
\maketitle
\tableofcontents{}

{}

{}

\section{Introduction}

In the past decade, the groundbreaking detections of gravitational
waves from binary black hole mergers by LIGO and Virgo have opened
up new avenues for testing general relativity in the strong field
regime \cite{Abbott:2016blz}. During the ringdown phase of a black
hole merger, the emitted gravitational waves can be described by a
superposition of quasinormal modes \cite{Berti:2007dg}. Extracting
these quasinormal modes from the observed gravitational wave signals
allows us to determine the properties of the remnant black hole, e.g.,
its mass, charge, and angular momentum, thus providing a promising
tool to test the validity of the Kerr hypothesis \cite{Price:2017cjr,Giesler:2019uxc}.
Furthermore, a significant breakthrough in the field of black hole
observation was achieved with the recent release of images of the
supermassive black holes M87{*} and Sgr A{*} by the Event Horizon
Telescope (EHT) collaboration \cite{Akiyama:2019bqs,Akiyama:2019brx,Akiyama:2019cqa,Akiyama:2019eap,Akiyama:2019fyp,Akiyama:2019sww,EventHorizonTelescope:2022exc,EventHorizonTelescope:2022urf,EventHorizonTelescope:2022vjs,EventHorizonTelescope:2022wok,EventHorizonTelescope:2022xnr,EventHorizonTelescope:2022xqj}.
While the detections of gravitational waves and the observed black
hole images align well with the predictions of Kerr black holes, it
is worth noting that the limitations in observational resolution leave
room for exploration of alternative theories and black hole mimics.

The no-hair theorem asserts that black holes are characterized exclusively
by their mass, electric charge and angular momentum \cite{Israel:1967wq,Carter:1971zc,Ruffini:1971bza}.
However, various models have been formulated that give rise to hairy
black holes endowed with additional degrees of freedom, thereby serving
as counterexamples to the no-hair theorem \cite{Volkov:1989fi,Bizon:1990sr,Greene:1992fw,Luckock:1986tr,Droz:1991cx,Kanti:1995vq,Mahapatra:2020wym,Priyadarshinee:2021rch,Priyadarshinee:2023cmi,Ghosh:2023kge}.
One prominent counterexample is spontaneous scalarization, which typically
occurs in models featuring non-minimal couplings between scalar fields
and other fields. These coupling terms serve as sources that destabilize
scalar-free black hole solutions and give rise to the formation of
scalarized hairy black holes.

Initially explored in the context of scalar-tensor models for neutron
stars, spontaneous scalarization was found when scalar fields are
coupled to the Ricci curvature \cite{Damour:1993hw}. It was demonstrated
that a coexistence region exists where scalar-free and scalarized
neutron star solutions compete energetically, with the scalarized
solutions often exhibiting a preference. Subsequently, it was discovered
that spontaneous scalarization can also manifest in black hole spacetime
in scalar-tensor models, provided that black holes are coupled to
non-linear electrodynamics \cite{Stefanov:2007eq,Doneva:2010ke} or
surrounded by non-conformally invariant matter \cite{Cardoso:2013opa,Cardoso:2013fwa}.
More recently, the phenomenon of spontaneous scalarization has been
investigated in extended Scalar-Tensor-Gauss-Bonnet (eSTGB) gravity
\cite{Doneva:2017bvd,Silva:2017uqg,Antoniou:2017acq,Doneva:2018rou,Cunha:2019dwb,Herdeiro:2020wei,Berti:2020kgk}.
In eSTGB models, the scalar field is non-minimally coupled to the
Gauss-Bonnet curvature correction in the gravitational sector, which
can induce the formation of spinning scalarized black holes \cite{Cunha:2019dwb,Herdeiro:2020wei}.
However, the presence of non-linear curvature terms in the eSTGB models
brings numerical challenges in solving the evolution equations.

To gain a deeper understanding of the dynamical evolution leading
to the formation of scalarized black holes, a technically simpler
class of models called Einstein-Maxwell-scalar (EMS) models has been
proposed \cite{Herdeiro:2018wub}. These models incorporate non-minimal
couplings between the scalar and Maxwell fields, which introduce tachyonic
instabilities capable of triggering spontaneous scalarization. In
\cite{Herdeiro:2018wub}, fully non-linear numerical simulations in
spherical symmetry demonstrated the evolution of Reissner-Nordström
(RN) black holes into scalarized RN black holes. Subsequent investigations
of the EMS models have yielded a wealth of research findings, e.g.,
different non-minimal coupling functions \cite{Fernandes:2019rez,Fernandes:2019kmh,Blazquez-Salcedo:2020nhs},
massive and self-interacting scalar fields \cite{Zou:2019bpt,Fernandes:2020gay},
horizonless reflecting stars \cite{Peng:2019cmm}, stability analysis
of scalarized black holes \cite{Myung:2018vug,Myung:2019oua,Zou:2020zxq,Myung:2020etf,Mai:2020sac},
higher dimensional scalar-tensor models \cite{Astefanesei:2020qxk},
quasinormal modes of scalarized black holes \cite{Myung:2018jvi,Blazquez-Salcedo:2020jee},
two U(1) fields \cite{Myung:2020dqt}, quasitopological electromagnetism
\cite{Myung:2020ctt}, topology and spacetime structure influences
\cite{Guo:2020zqm}, scalarized black hole solutions in the dS/AdS
spacetime \cite{Brihaye:2019dck,Brihaye:2019gla,Zhang:2021etr,Guo:2021zed,Chen:2023eru},
and dynamical scalarization and descalarization \cite{Zhang:2021nnn,Zhang:2022cmu,Jiang:2023yyn}.

Remarkably, scalarized RN black holes have been discovered to exhibit
the presence of multiple photon spheres outside the event horizon
in specific parameter regimes \cite{Gan:2021pwu}. Subsequent investigations
have focused on the optical appearances of various phenomena in the
background of scalarized RN black holes, e.g., accretion disks \cite{Gan:2021pwu,Gan:2021xdl},
luminous celestial spheres \cite{Guo:2022muy} and infalling stars
\cite{Chen:2022qrw}. These studies have revealed that the existence
of an additional photon sphere significantly increases the flux of
observed accretion disk images, generates triple higher-order images
of a luminous celestial sphere, and gives rise to an additional cascade
of flashes from an infalling star. Furthermore, the presence of multiple
photon spheres in a spacetime also suggests the existence of long-lived
modes that may render the spacetime unstable \cite{Cardoso:2014sna,Keir:2014oka,Guo:2021bcw,Guo:2021enm,Guo:2022umh}.
Specifically, it has been shown that the existence of multiple photon
spheres outside the event horizon can induce superradiance instabilities
for charged scalar perturbations \cite{Guo:2023ivz}. For a more detailed
analysis of black holes with multiple photon spheres, we refer readers
to the work \cite{Guo:2022ghl}.

Recently, it has been reported that tachyonic instabilities around
Kerr-Newman (KN) black holes can lead to unstable scalar perturbations
in the EMS model, indicating the existence of scalarized rotating
black holes \cite{Lai:2022ppn}. The aim of this paper is twofold:
first, to numerically construct scalarized rotating black hole solutions
in the EMS model; and second, to investigate whether scalarized rotating
black holes possess multiple light rings. The rest of the paper is
organized as follows. In Section \ref{sec:EMS Model}, we introduce
the construction of scalarized KN black holes in the EMS model and
discuss light rings on the equatorial plane. Subsequently, we present
the numerical results for the scalarized black hole solutions and
their light ring structure in Section \ref{sec:NR}. We conclude our
main results and provide discussions in Section \ref{Sec:Conc}. We
set $G=c=4\pi\epsilon_{0}=1$ throughout this paper.

\section{Einstein-Maxwell-scalar Model}

\label{sec:EMS Model}

In the EMS model, where a scalar field is non-minimally coupled to
electromagnetism, the background spacetime can become destabilized
through a tachyonic instability, leading to the phenomenon of spontaneous
scalarization in black holes. The corresponding action is described
as 
\begin{equation}
S=\frac{1}{16\pi}\int d^{4}x\sqrt{-g}\left[R-2\partial_{\mu}\phi\partial^{\mu}\phi-f\left(\phi\right)F^{\mu\nu}F_{\mu\nu}\right],\label{eq:Action}
\end{equation}
where $\phi$ is the scalar field, $F_{\mu\nu}=\partial_{\mu}A_{\nu}-\partial_{\nu}A_{\mu}$
denotes the electromagnetic field strength tensor, and $A_{\mu}$
represents the electromagnetic field. It is important to note that
to trigger spontaneous scalarization, the coupling function $f\left(\phi\right)$
between $\phi$ and $A_{\mu}$ should satisfy $f\left(0\right)=1$
and $f^{\prime}\left(0\right)\equiv\left.df\left(\phi\right)/d\phi\right\vert _{\phi=0}=0$
\cite{Herdeiro:2018wub,Fernandes:2019rez}. In this paper, we focus
on the specific exponential coupling function $f\left(\phi\right)=e^{\alpha\phi^{2}}$
with $\alpha>0$. In this section, we present the numerical construction
of scalarized KN black holes in the EMS model and investigate null
circular geodesics (i.e., light rings) within these scalarized KN
black holes.

\subsection{Tachyonic Instability}

In the scalar-free background, represented by KN black holes, the
scalar perturbation $\delta\phi$ is governed by the equation 
\begin{equation}
\left(\square-\mu_{\text{eff}}^{2}\right)\delta\phi=0,\label{eq:delta phi}
\end{equation}
where $\mu_{\text{eff}}^{2}=\alpha F^{\mu\nu}F_{\mu\nu}/2$. In the
Boyer-Linquist coordinates, the effective mass square can be expressed
as 
\begin{equation}
\mu_{\text{eff}}^{2}=-\frac{\alpha Q^{2}\left(r^{4}-6a^{2}r^{2}\cos^{2}\theta+a^{4}\cos^{4}\theta\right)}{\left(r^{2}+a^{2}\cos^{2}\theta\right)^{4}},\label{eq:effmass}
\end{equation}
where $Q$ is the black hole charge, and $a$ represents the ratio
of black hole angular momentum $J$ to mass $M$ (i.e., $a\equiv J/M$).
Notably, tachyonic instabilities arise in the presence of a negative
effective mass square $\mu_{\text{eff}}^{2}<0$, which could induce
scalarized black holes from the scalar-free background. When restricted
to the equatorial plane, the effective mass square reduces to $\mu_{\text{eff}}^{2}=-\alpha Q^{2}/r^{4}$,
which coincides with the RN black hole case \cite{Herdeiro:2018wub,Guo:2021zed}.
In the RN black hole background, it has demonstrated that the tachyonic
instabilities near the black hole event horizon can be strong enough
to trigger spontaneous scalarization. By analogy, we expect similar
tachyonic instabilities to induce scalarized KN black holes from the
KN black hole background.

To investigate the onset of spontaneous scalarization in KN black
holes, we need to solve the linear perturbation equation $\left(\ref{eq:delta phi}\right)$
for zero-modes $\delta\phi\left(r,\theta\right)$. The presence of
these zero-modes often indicates the existence of scalarized black
hole solutions. Specifically, the zero-modes correspond to bifurcation
points in the parameter space, marking the onset of scalarized KN
black holes. In the latter part of this section, we will discuss the
numerical scheme employed to find the zero-modes $\delta\phi\left(r,\theta\right)$.

\subsection{Rotating Black Hole Solution}

The equations of motion for the metric field $g_{\mu\nu}$, the scalar
field $\phi$ and the electromagnetic field $A_{\mu}$ can be obtained
by varying the action $\left(\ref{eq:Action}\right)$, yielding 
\begin{align}
R_{\mu\nu}-\frac{1}{2}Rg_{\mu\nu} & =2T_{\mu\nu},\nonumber \\
\square\phi-\frac{\alpha}{2}\phi e^{\alpha\phi^{2}}F^{\mu\nu}F_{\mu\nu} & =0,\label{eq:nonlinearEOMs}\\
\partial_{\mu}\left(\sqrt{-g}e^{\alpha\phi^{2}}F^{\mu\nu}\right) & =0,\nonumber 
\end{align}
where the energy-momentum tensor $T_{\mu\nu}$ is expressed as 
\begin{equation}
T_{\mu\nu}=\partial_{\mu}\phi\partial_{\nu}\phi-\frac{1}{2}g_{\mu\nu}\left(\partial\phi\right)^{2}+e^{\alpha\phi^{2}}\left(F_{\mu\rho}F_{\nu}^{\text{ }\rho}-\frac{1}{4}g_{\mu\nu}F_{\rho\sigma}F^{\rho\sigma}\right).
\end{equation}
Following \cite{Herdeiro:2015gia,Delgado:2016jxq,Herdeiro:2020wei},
we consider stationary, axisymmetric and asymptotically-flat black
hole solutions with the generic ansatz 
\begin{align}
ds^{2} & =-e^{2F_{0}}Ndt^{2}+e^{2F_{1}}\left(\frac{dr^{2}}{N}+r^{2}d\theta^{2}\right)+e^{2F_{2}}r^{2}\sin^{2}\theta\left(d\varphi^{2}-\frac{W}{r^{2}}dt\right)^{2},\nonumber \\
A_{\mu}dx^{\mu} & =\left(A_{t}-A_{\varphi}\frac{W}{r^{2}}\sin\theta\right)dt+A_{\varphi}\sin\theta d\varphi\text{ and }\phi=\phi\left(r,\theta\right),\label{eq:ansatz}
\end{align}
where $N\equiv1-1/r_{H}$, and $r_{H}$ is the black hole horizon.
Here, the metric functions $F_{0}$, $F_{1}$, $F_{2}$, $W$, $A_{t}$
and $A_{\varphi}$ are assumed to depend solely on the coordinates
$r$ and $\theta$. By substituting the ansatz $\left(\ref{eq:ansatz}\right)$
into eqn. $\left(\ref{eq:nonlinearEOMs}\right)$, we derive a set
of nonlinear partial differential equations for the metric functions.
The solutions of the partial differential equations can describe scalarized
KN black holes with a nontrivial profile of the scalar field or KN
black holes with $\phi=0$.

In the stationary spacetime, two Killing vectors $\partial_{t}$ and
$\partial_{\varphi}$ are present, and their combination $\xi=\partial_{t}+\Omega_{H}\partial\varphi$,
where $\Omega_{H}$ is the angular velocity of the black hole horizon,
is both orthogonal to and null at the horizon. Consequently, the surface
gravity $\kappa$ is defined as $\kappa^{2}=-\left(\nabla_{\mu}\xi_{\nu}\right)\left(\nabla^{\mu}\xi^{\nu}\right)/2$,
and it is related to the Hawking temperature $T_{H}$ as given in
\cite{Herdeiro:2015gia} 
\begin{equation}
T_{H}=\frac{\kappa}{2\pi}=\frac{1}{4\pi r_{H}}e^{F_{0}\left(r_{H},\theta\right)-F_{1}\left(r_{H},\theta\right)}.\label{eq:TH}
\end{equation}
In the EMS model, the black hole entropy is expressed as $S=A_{H}/4$,
where the area of the horizon $A_{H}$ is given by 
\begin{equation}
A_{H}=2\pi r_{H}^{2}\int_{0}^{\pi}d\theta\sin\theta e^{F_{1}\left(r_{H},\theta\right)+F_{2}\left(r_{H},\theta\right)}.\label{eq:Ah}
\end{equation}

Various physical quantities, such as the black hole mass $M$, the
black hole charge $Q$, the black hole angular momentum $J$, the
electrostatic potential $\Phi$ and the horizon angular velocity $\Omega_{H}$,
can be determined by studying the asymptotic behavior of the metric
functions at the event horizon and spatial infinity \cite{Herdeiro:2015gia,Delgado:2016jxq},
\begin{align}
\left.A_{t}\right\vert _{r=r_{H}} & \sim0,\quad\left.W\right\vert _{r=r_{H}}\sim r_{H}^{2}\Omega_{H},\nonumber \\
\left.A_{t}\right\vert _{r=\infty} & \sim\Phi-\frac{Q}{r},\quad\left.W\right\vert _{r=\infty}\sim\frac{2J}{r},\quad\left.e^{2F_{0}}N\right\vert _{r=\infty}\sim1-\frac{2M}{r}.\label{eq:asymptotic behaviors}
\end{align}
Moreover, these physical quantities are related by the Smarr relation
\cite{Herdeiro:2015gia,Guo:2021zed,Fernandes:2022gde} 
\begin{equation}
M=2T_{H}S+2\Omega_{H}J+\Phi Q,\label{eq:smarr}
\end{equation}
which enables us to assess the accuracy of our numerical black hole
solutions.

\subsection{Light Ring}

Light rings are null circular geodesics in the black hole spacetime
and plays a crucial role in strong gravitational lensing and in the
formation of black hole images. For simplicity, we focus on light
rings on the equatorial plane. On the equatorial plane, the motion
of photons is described by the Lagrangian 
\begin{equation}
\mathcal{L}=-\frac{e^{2F_{0}}N\dot{t}^{2}}{2}+\frac{e^{2F_{1}}}{2N}\dot{r}^{2}+\frac{e^{2F_{2}}r^{2}}{2}\left(\dot{\phi}-\frac{W}{r^{2}}\dot{t}\right)^{2},\label{eq:photon Lag}
\end{equation}
where dots denote derivatives with respect to an affine parameter
$\lambda$. The conserved energy $E$ and angular momentum $L$ of
photons, associated with the Killing vectors $\partial_{t}$ and $\partial_{\varphi}$,
respectively, are expressed as 
\begin{align}
E & =\left(e^{2F_{0}}N-e^{2F_{2}}\frac{W^{2}}{r^{2}}\right)\dot{t}+e^{2F_{2}}W\dot{\varphi},\nonumber \\
L & =e^{2F_{2}}\left(r^{2}\dot{\phi}-W\dot{t}\right).\label{eq:conserved quantitites}
\end{align}
By substituting eqn. $\left(\ref{eq:conserved quantitites}\right)$
into eqn. $\left(\ref{eq:photon Lag}\right)$, the constancy of the
Lagrangian $\mathcal{L}=0$ reduces to the radial equation for photons,
\begin{equation}
\dot{r}^{2}+V_{\text{eff}}=0\text{,}
\end{equation}
where $V_{\text{eff}}\leq0$ for null geodesics. Here, the effective
potential is defined as 
\begin{equation}
V_{\text{eff}}=L^{2}e^{-2F_{1}}N\left[\frac{e^{-2F_{2}}}{r^{2}}-\frac{e^{-2F_{0}}}{N}\left(\frac{1}{b}-\frac{W}{r^{2}}\right)^{2}\right],\label{eq:effV}
\end{equation}
where $b\equiv L/E$ is the impact parameter. Thus, for a photon with
an impact parameter $b_{c}$, it would follow a circular orbit at
$r=r_{c}$ if $V_{\text{eff}}\left(b_{c},r_{c}\right)=0$ and $\partial_{r}V_{\text{eff}}\left(b_{c},r_{c}\right)=0$.

To facilitate further analysis, we can factorize the effective potential
$V_{\text{eff}}$ as shown in \cite{Cunha:2016bjh},
\begin{equation}
V_{\text{eff}}=-L^{2}e^{-2F_{1}-2F_{0}}\left(\frac{1}{b}-H_{+}\right)\left(\frac{1}{b}-H_{-}\right),
\end{equation}
where 
\begin{equation}
H_{+}=\frac{\sqrt{e^{2F_{0}-2F_{2}}Nr^{2}}+W}{r^{2}},\quad H_{-}=-\frac{\sqrt{e^{2F_{0}-2F_{2}}Nr^{2}}-W}{r^{2}}.\label{eq:H+-}
\end{equation}
Notably, the two redefined effective potentials, $H_{+}$ and $H_{-}$,
are independent of the impact parameter and describe prograde and
retrograde motion of photons on the equatorial plane, respectively.
For a null geodesic, the effective potentials $H_{+}$ and $H_{-}$
must satisfy $H_{+}<b^{-1}$ or $H_{-}>b^{-1}$, where $H_{+}\geq H_{-}$.
In particular, a null circular geodesic at $r=r_{c}$ corresponds
to a local extremum of $H_{+}$ or $H_{-}$, meaning $\partial_{r}H_{+}\left(r_{c}\right)=0$
or $\partial_{r}H_{-}\left(r_{c}\right)=0$. Moreover, $\partial_{r}^{2}H_{+}\left(r_{c}\right)>0$
and $\partial_{r}^{2}H_{+}\left(r_{c}\right)<0$ indicate stable and
unstable prograde light rings, respectively, while $\partial_{r}^{2}H_{-}\left(r_{c}\right)<0$
and $\partial_{r}^{2}H_{-}\left(r_{c}\right)>0$ correspond to stable
and unstable retrograde light rings, respectively. It is worth mentioning
that in the static case with $W=0$, light rings can be solely determined
by either $H_{+}$ or $H_{-}$ since $H_{+}=-H_{-}$.

\subsection{Numerical Scheme}

In this paper, we utilize pseudospectral methods to numerically solve
the linear partial differential equation $\left(\ref{eq:delta phi}\right)$
for zero-modes $\delta\phi\left(r,\theta\right)$ and a set of coupled
nonlinear partial differential equations for scalarized KN black holes.
Pseudospectral methods are a well-established approach for solving
partial differential equations \cite{boyd2001chebyshev}. They approximate
the exact solution by a finite linear combination of basis functions.
Notably, as the number of degrees-of-freedom increases, pseudospectral
methods exhibit an exponential convergence rate for well-behaved functions,
in contrast to the linear or polynomial convergence of finite difference
or finite element methods. Recently, these pseudospectral methods
have demonstrated successful applications in searching for black hole
solutions \cite{Fernandes:2022gde,Lai:2023gwe,Burrage:2023zvk}, as
well as computing black hole quasinormal modes \cite{Jansen:2017oag,Gan:2019jac,Chung:2023zdq}.
For technical details of pseudospectral methods in the context of
black hole physics, interested readers can refer to \cite{Fernandes:2022gde}.

For the numerical implementation, we introduce a new radial variable
given by 
\begin{equation}
x=\frac{\sqrt{r^{2}-r_{H}^{2}}-r_{H}}{\sqrt{r^{2}-r_{H}^{2}}+r_{H}},
\end{equation}
which allows us to map the event horizon and spatial infinity to $x=-1$
and $x=1$, respectively. By performing series expansions of the solutions
at the event horizon, we obtain the corresponding boundary conditions
\begin{equation}
\partial_{x}\delta\phi=\partial_{x}F_{0}=\partial_{x}F_{1}=\partial_{x}F_{2}=\partial_{x}\phi=\partial_{x}A_{\varphi}=A_{t}=W-\Omega_{H}=0\text{ at }x=-1.\label{eq:bdx0}
\end{equation}
Moreover, due to the flatness at the spatial infinity, we have 
\begin{equation}
\delta\phi=F_{0}=F_{1}=F_{2}=\phi=A_{\varphi}=A_{t}-\Phi=W=0\text{ at }x=1.\label{eq:bdx1}
\end{equation}
On the other hand, the regularity on the symmetric axis imposes the
conditions 
\begin{equation}
\partial_{\theta}\delta\phi=\partial_{\theta}F_{0}=\partial_{\theta}F_{1}=\partial_{\theta}F_{2}=\partial_{\theta}\phi=\partial_{\theta}A_{\varphi}=\partial_{\theta}A_{t}=\partial_{\theta}W=0\text{ at }\theta=0\,\text{and }\pi.\label{eq:bdtheta0}
\end{equation}
Since all solutions are symmetric with respect to the equatorial plane,
we can restrict the analysis to the upper half domain with $0\leq\theta\leq\pi/2$,
and hence replace the $\theta=\pi$ boundary condition in eqn. $\left(\ref{eq:bdtheta0}\right)$
with 
\begin{equation}
\partial_{\theta}\delta\phi=\partial_{\theta}F_{0}=\partial_{\theta}F_{1}=\partial_{\theta}F_{2}=\partial_{\theta}\phi=\partial_{\theta}A_{\varphi}=\partial_{\theta}A_{t}=\partial_{\theta}W=0\text{ at }\theta=\pi/2.\label{eq:bdtheta1}
\end{equation}
Consequently, eqns. $\left(\ref{eq:bdx0}\right)$, $\left(\ref{eq:bdx1}\right)$,
$\left(\ref{eq:bdtheta0}\right)$ and $\left(\ref{eq:bdtheta1}\right)$
serve as boundary conditions for solving the partial differential
equations. Additionally, the absence of conical singularities requires
$F_{1}=F_{2}$ on the symmetric axis, which can be used to verify
our numerical results as well as the Smarr relation \cite{Herdeiro:2015gia,Herdeiro:2020wei}.

With the compactified radial coordinate $x$, the functions of interest,
which are collectively denoted by $\mathcal{F}=\left\{ F_{0},F_{1},F_{2},W,A_{t},A_{\varphi},\phi,\delta\phi\right\} $,
can be decomposed into a spectral expansion,
\begin{equation}
\mathcal{F}^{\left(k\right)}=
{\displaystyle \sum\limits _{i=0}^{N_{x}-1}}
{\displaystyle \sum\limits _{j=0}^{N_{\theta}-1}}
\alpha_{ij}^{\left(k\right)}T_{i}\left(x\right)\cos\left(2j\theta\right).\label{eq:sexpansion}
\end{equation}
Here, $N_{x}$ and $N_{\theta}$ represent the resolutions in the
radial and angular coordinates, respectively, $T_{i}\left(x\right)$
is the Chebyshev polynomial, and $\alpha_{ij}^{\left(k\right)}$ are
the spectral coefficients. To ensure numerical precision and efficiency,
we set $\left(N_{x},N_{\theta}\right)=\left(22,5\right)$ and $\left(N_{x},N_{\theta}\right)=\left(42,8\right)$
for the subsequent numerical computations of the zero-modes and the
metric functions, respectively. To determine $\alpha_{ij}^{\left(k\right)}$,
we substitute the spectral expansions $\left(\ref{eq:sexpansion}\right)$
into the partial differential equations and discretize the resulting
equations at the Gauss-Chebyshev points. This process reduces the
partial differential equations of $\mathcal{F}$ to a finite system
of algebraic equations of $\alpha_{ij}$. We then solve these algebraic
equations for $\alpha_{ij}$ by a standard iterative Newton-Raphson
method, where the resulting linear system of equations is solved using
the built-in command LinearSolve in Mathematica.

\section{Numerical Results}

\label{sec:NR}

In this section, we explore the domain of existence for scalarized
KN black hole solutions and conduct an analysis of the light ring
structure. To facilitate our study, we introduce dimensionless reduced
quantities, $q\equiv Q/M$, $\chi\equiv a/M$ and $a_{H}\equiv A_{H}/16\pi M^{2}$.
Throughout the numerical implementation, we carefully assess the accuracy
of the black hole solutions by verifying the absence of conical singularities
and the Smarr relation. Our results reveal that the scalarized KN
black hole solutions exhibit a numerical error on the order of $10^{-8}$.

\subsection{Domain of Existence}

\begin{figure}[ptb]
\begin{centering}
\includegraphics[scale=0.71]{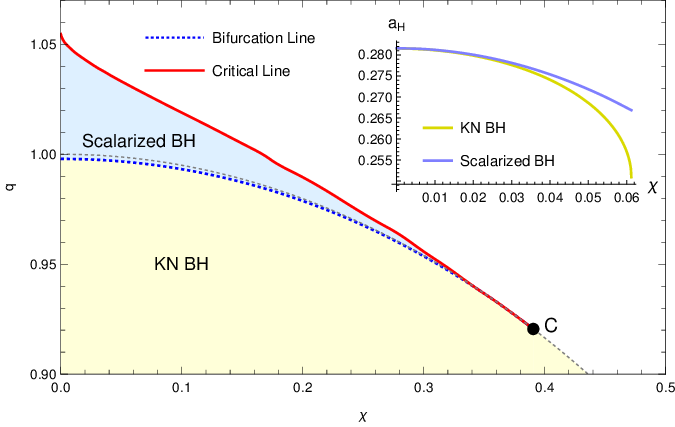} \includegraphics[scale=0.71]{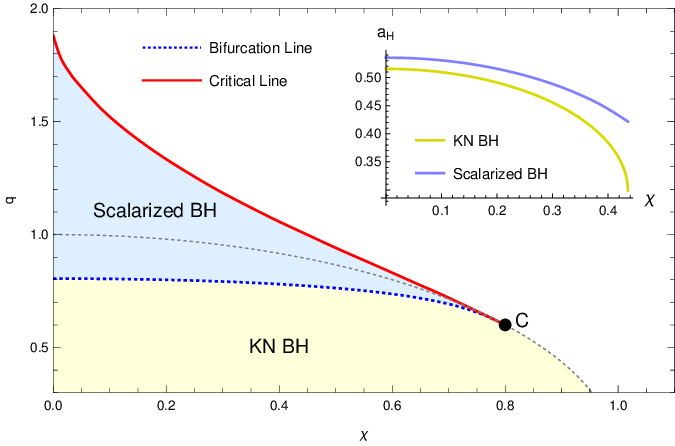} 
\par\end{centering}
\caption{Domain of existence for scalarized KN black holes in the $q$-$\chi$
plane, with $\alpha=0.8$ (\textbf{left}) and $\alpha=5$ (\textbf{right}),
depicted by the regions shaded in light blue. The domain of existence
is bounded by the bifurcation and critical lines. The blue dashed
lines represent the bifurcation lines, indicating the points where
scalarized black holes emerge from KN black holes as zero modes. Meanwhile,
the red lines indicate critical configurations of scalarized black
holes, characterized by a vanishing horizon area while the mass and
charge remain finite. Interestingly, as the black hole spin increases,
the bifurcation and critical lines approach each other and eventually
converge at the critical point $C$, illustrating the suppressive
effect of rotation on scalarization. KN black holes are situated below
the extreme lines, marked by dashed gray lines, and coexist with scalarized
black holes in the regions lying between the bifurcation and extreme
lines. The insets in the left and right panels depict the reduced
horizon area as a function of $\chi$ for $q=0.998$ scalarized black
holes with $\alpha=0.8$ and $q=0.900$ scalarized black holes with
$\alpha=5$, respectively. These insets demonstrate that scalarized
black holes are entropically favored in the coexistence regions.}
\label{ExistenceDomain}
\end{figure}

To determine the existence domain of scalarized KN black holes, we
adopt a step-by-step approach. Initially, we obtain scalarized black
hole solutions along the extreme line of KN black holes by iteratively
using the solution from the previous step as the starting values for
Newton-Raphson computations at the current step. This procedure starts
with the scalarized RN black hole solution, where $q=1$, and gradually
leads to scalarized KN black hole solutions on the extreme line. Subsequently,
we proceed by varying $q$ to compute scalarized black hole solutions
along $\chi$-cosntant lines until it becomes unfeasible to find additional
solutions.

In Fig. \ref{ExistenceDomain}, the left and right panels present
the existence domain of scalarized KN black holes with $\alpha=0.8$
and $\alpha=5$, respectively, in the $q$-$\chi$ parameter space.
The light blue region represents the domain where scalarized black
holes exist, and it is delimited by the bifurcation line and the critical
line. The bifurcation lines indicate the threshold where the tachyonic
instabilities near the event horizon of scalar-free black holes are
strong enough to trigger the formation of scalarized black holes.
Remarkably, our numerical findings demonstrate that the bifurcation
lines coincide perfectly with the zero-mode lines in the $q$-$\chi$
parameter space. This coincidence is expected since scalarized black
hole solutions originate from the bifurcation lines as zero modes.
On the other hand, scalarized black hole solutions lying on the critical
lines possess a vanishing horizon area, while the black hole mass
$M$ and charge $Q$ remain finite.

Fig. \ref{ExistenceDomain} demonstrates the occurrence of spontaneous
scalarization in the EMS model for rotating black holes; however,
it is suppressed for high spins. Notably, the bifurcation and critical
lines converge and terminate at the critical point $C$ on the extreme
line, setting an upper limit on the spin of scalarized black holes.
This observation suggests that sufficiently large spin values mitigate
tachyonic instabilities, thereby impeding the initiation of spontaneous
scalarizations. Moreover, a larger coupling constant $\alpha$ facilitates
the scalarization of KN black holes, leading to an expansion of the
existence domain for scalarized black holes.

Of particular interest is the coexistence region where scalarized
and KN black holes both exist, and this region is delimited by the
bifurcation and extreme lines. The inset in each panel displays that
scalarized black holes consistently exhibit a greater reduced area
of the event horizon when compared to KN black holes. This compelling
observation suggests that within the coexistence region, scalarized
black holes are entropically favored over scalar-free black holes.
Lastly, it is worth noting that scalarized black holes can be overcharged
in the parameter region lying between the extremal and critical lines.

\subsection{Light Ring Structure}

\label{subsec:Light-Ring}

\begin{figure}[t]
\begin{centering}
\includegraphics[scale=0.6]{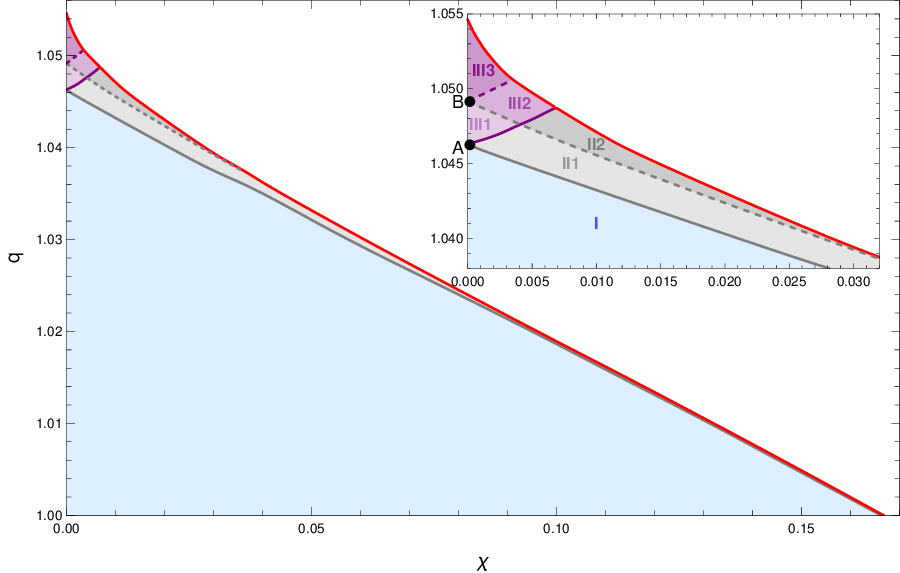} \includegraphics[scale=0.4]{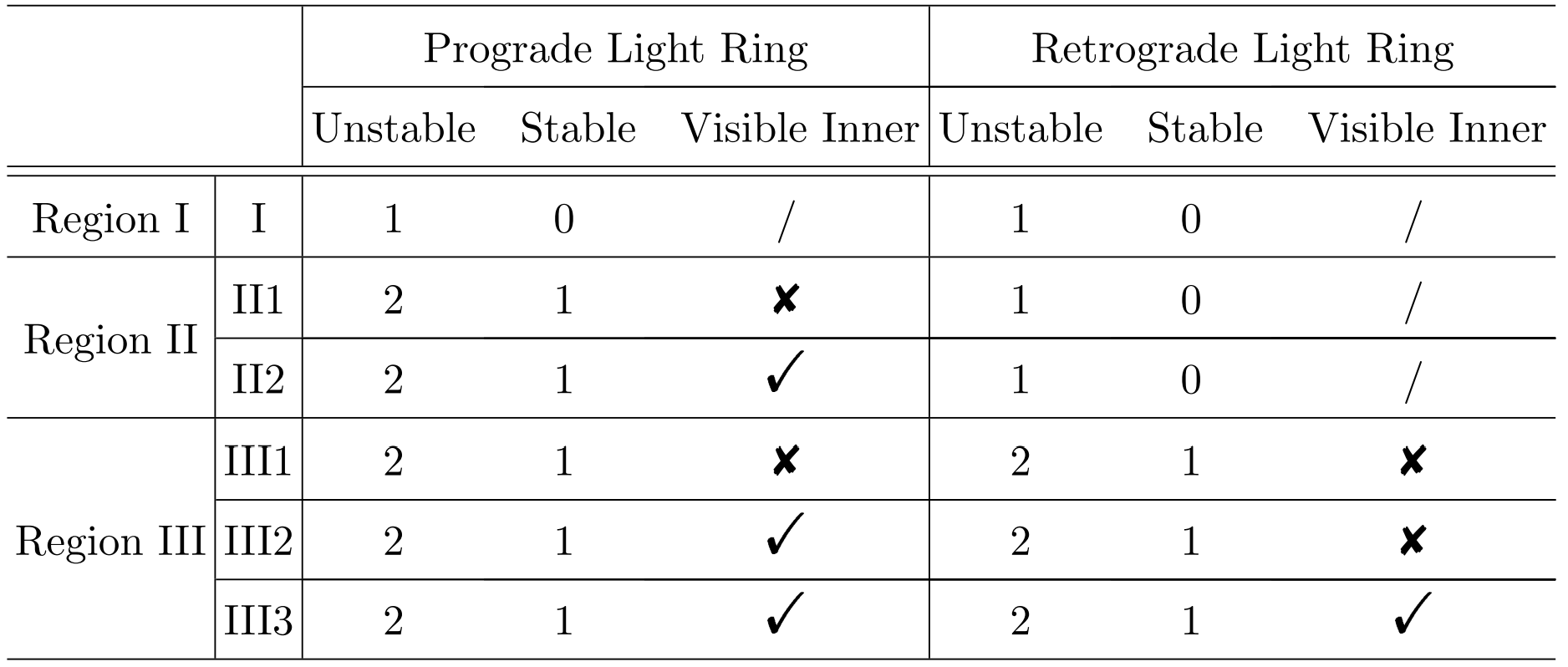} 
\par\end{centering}
\caption{\textbf{Upper}: Scalarized black holes exhibit distinct light ring
structures in various regions of the $q$-$\chi$ plane. The boundaries
separating these regions are represented by threshold curves, and
the corresponding effective potentials are illustrated in FIG. \ref{fig:thresholdLines}.
Here, we set $\alpha=0.8$. \textbf{Lower}: The table displays the
number of unstable and stable light rings for scalarized black holes
in the six regions as shown in the upper panel. The table header \textquotedblleft Visible
Inner\textquotedblright\ indicates whether photons originating from
inner light rings can reach a distant observer.}
\label{LRpara}
\end{figure}

\begin{figure}[t]
\begin{centering}
\includegraphics[scale=0.7]{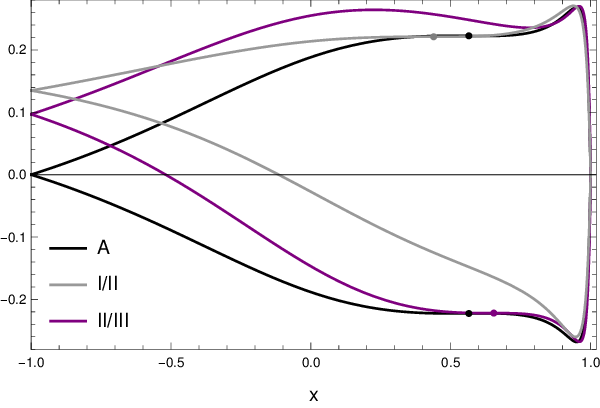} \hspace{5pt} \includegraphics[scale=0.7]{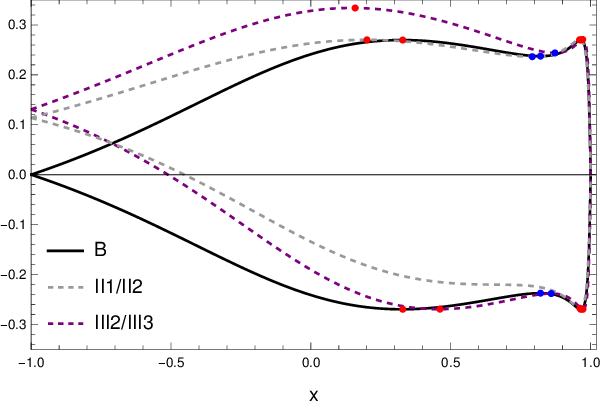} 
\par\end{centering}
\caption{The effective potentials as a function of $x$ for scalarized black
holes with $\alpha=0.8$ on the threshold curves. The upper and lower
branches of the effective potentials denotes the prograde potential
$H_{+}$ and the retrograde potential $H_{-}$, respectively. \textbf{Left}:
The effective potentials are displayed for the black hole with $q=1.0463$
and $\chi=0$ at point $A$, the one with $q=1.0400$ and $\chi=0.0202$
on the I/II curve, and the one with $q=1.0475$ and $\chi=0.0038$
on the II/III curve. Inflection points are marked with colored dots,
which transform into a maximum and a minimum as $q$ increases. \textbf{Right}:
The effective potentials are shown for the black hole with $q=1.0492$
and $\chi=0$ at point $B$, the one with $q=1.0475$ and $\chi=0.0044$
on the II1/II2 curve, and the one with $q=1.0500$ and $\chi=0.0021$
on the III2/III3 curve. Unstable and stable light rings are represented
by red and blue dots, respectively. The potential peaks or wells at
two unstable light rings possess identical height or depth for $H_{\pm}$
at point $B$, $H_{+}$ on the II1/II2 curve and $H_{-}$ on the III2/III3
curve. }
\label{fig:thresholdLines}
\end{figure}

\begin{figure}[t]
\begin{centering}
\includegraphics[scale=0.7]{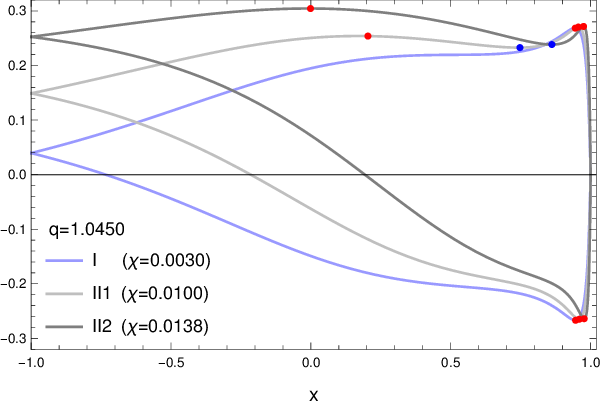} \hspace{5pt} \includegraphics[scale=0.7]{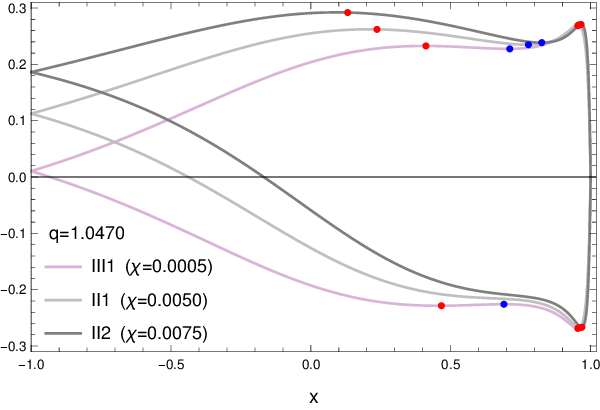}
\includegraphics[scale=0.7]{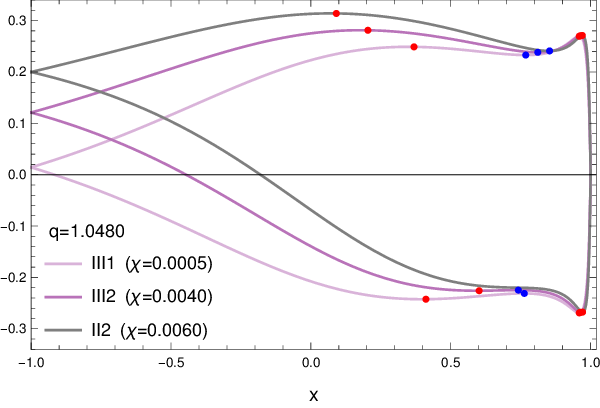} \hspace{5pt} \includegraphics[scale=0.7]{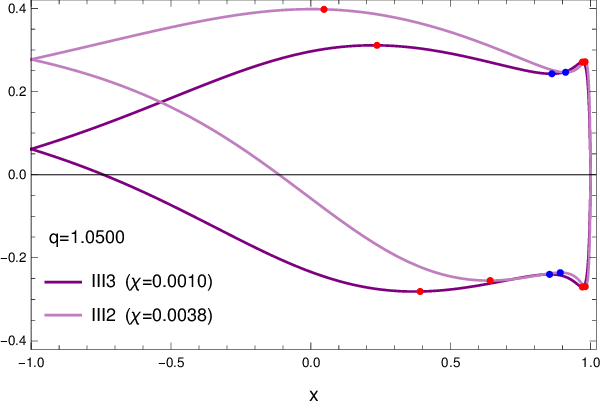} 
\par\end{centering}
\caption{The effective potentials as a function of $x$ for representative
scalarized black hole solutions with $\alpha=0.8$ in the regions
characterized by distinct light ring structures. The upper and lower
branches of the effective potentials are denoted as the prograde potential
$H_{+}$ and the retrograde potential $H_{-}$, respectively. Red
and blue dots represent unstable and stable light rings, respectively. }
\label{Heffs}
\end{figure}

As discussed earlier, the maxima and minima of $H_{+}$ correspond
to unstable and stable prograde light rings, respectively, while the
maxima and minima of $H_{-}$ correspond to stable and unstable retrograde
light rings, respectively. To study the light ring structure on the
equatorial plane, we investigate the extrema of the effective potentials
$H_{+}$ and $H_{-}$ for scalarized KN black holes. In the upper
panel of FIG. \ref{LRpara}, we present the light ring structure for
scalarized black holes with $\alpha=0.8$ in the $q$-$\chi$ plane,
with a zoomed-in inset providing more details. When $\chi=0$, the
threshold points $A$ with $q_{A}=1.0463$ and $B$ with $q_{B}=1.0492$
determine the structure of light rings (or equivalently, photon spheres
in the static case) for scalarized RN black holes. The black lines
in the left and right panels of FIG. \ref{fig:thresholdLines} represent
the effective potentials of scalarized RN black holes at points $A$
and $B$, respectively. At the threshold point $A$, the effective
potential $H_{+}$ $\left(H_{-}\right)$ displays a maximum (minimum)
and an inflection point, which splits into a minimum (maximum) and
an additional maximum (minimum) for $q>q_{A}$. Consequently, scalarized
RN black holes with $q>q_{A}$ possess two unstable and one stable
light rings. On the other hand, at the threshold point $B$, there
are two potential peaks (well) of the effective potential $H_{+}$
$\left(H_{-}\right)$ with the same height (depth). For $q>q_{B}$,
the potential peak (well) at the inner light ring is higher (lower)
than the one at the outer ring, indicating that the inner light ring
is visible to a distant observer, and thus plays a significant role
in determining the optical appearances of luminous matters \cite{Gan:2021xdl,Gan:2021pwu,Guo:2022muy,Chen:2022qrw}.
Accordingly, light rings at the higher (lower) inner potential peak
(well) of $H_{+}$ $\left(H_{-}\right)$ are referred to as visible
inner light rings hereinafter.

As scalarized black holes acquire spin, a single threshold point transforms
into two threshold curves, each associated with prograde and retrograde
light rings, respectively. In FIG. \ref{LRpara}, the resulting four
threshold curves are as follows:
\begin{itemize}
\item The I/II curve, represented by the solid gray line, exhibits a maximum
and an inflection point in $H_{+}$. The left panel of FIG. \ref{fig:thresholdLines}
provides a typical example of the effective potentials, depicted by
the solid gray lines.
\item The II/III curve, shown by the solid purple line, features a minimum
and an inflection point in $H_{-}$. The left panel of FIG. \ref{fig:thresholdLines}
illustrates a typical example of the effective potentials, denoted
by the solid purple lines.
\item The II1/II2 curve, denoted by the dashed gray line, has two potential
peaks of $H_{+}$ at the same height. The right panel of FIG. \ref{fig:thresholdLines}
displays a typical example of the effective potentials, represented
by the dashed grays.
\item The III2/III3 curve, indicated by the dashed purple line, involves
two potential wells of $H_{-}$ with the same depth. The right panel
of FIG. \ref{fig:thresholdLines} shows a typical example of the effective
potentials, depicted by the dashed purple lines. 
\end{itemize}
As a consequence, the four threshold curves partition the domain of
existence into six regions, each corresponding to a distinct light
ring structure, as depicted in the upper panel of FIG. \ref{LRpara}.
The light ring properties in these six regions are presented in the
lower table of FIG. \ref{LRpara}. Additionally, we showcase the effective
potentials of representative black hole solutions in FIG. \ref{Heffs}.
Remarkably, our findings indicate that the number of unstable light
rings is always one greater than the number of stable ones, consistent
with prior observations in \cite{Cunha:2020azh}. Intriguingly, for
a given $q$, a sufficiently high spin can engender multiple light
rings, even from scalarized RN black holes with only one light ring.
Lastly, it is worth noting that prograde photons are more likely to
possess multiple light rings compared to retrograde photons.

\section{Conclusions}

\label{Sec:Conc}

This paper first presented our numerical construction of scalarized
rotating black hole solutions within the EMS model and explored the
parameter regions in which these solutions exist. The EMS model incorporates
a non-minimal coupling between the scalar and electromagnetic fields,
leading to tachyonic instabilities that can trigger the formation
of scalarized KN black holes. Our findings demonstrated that the black
hole spin tends to inhibit the spontaneous scalarization of KN black
holes, and scalarized KN black holes cease to exist beyond a certain
threshold spin. Later, we delved into the light ring structure of
scalarized KN black holes. We found that slowly-rotating scalarized
black holes with a small electric charge typically possess one unstable
prograde and one unstable retrograde light rings. However, as scalarized
black holes spin faster, two unstable and one stable prograde light
rings can emerge. Additionally, for sufficiently large black hole
charge, two unstable and one stable retrograde light rings can also
be observed.

While our focus has been on positive coupling constants $\alpha$,
it is noteworthy that studies involving negative coupling constants
$\alpha$ have revealed tachyonic instabilities leading to unstable
perturbations for large black hole spin \cite{Lai:2022spn,Hod:2022txa}.
Investigating the existence of scalarized black holes with negative
coupling constants $\alpha$ and understanding their formation from
scalar-free black holes are promising avenues for future research.
Furthermore, this work has shed light on the intricate behavior of
scalarized rotating black holes within the EMS model. Future studies
may explore the stability of these solutions and investigate their
observable signatures, taking advantage of the rich light ring structure
they exhibit. Such investigations will further enhance our understanding
of rotating black holes with multiple light rings and contribute to
the broader field of black hole physics.
\begin{acknowledgments}
We are grateful to Yiqian Chen and Qingyu Gan for useful discussions
and valuable comments. This work is supported in part by NSFC (Grant
No. 12105191, 12275183, 12275184 and 11875196). Houwen Wu is supported
by the International Visiting Program for Excellent Young Scholars
of Sichuan University. 
\end{acknowledgments}

 \bibliographystyle{unsrturl}
\bibliography{ref}

\end{document}